\documentclass[preprint,12pt]{elsarticle}

\usepackage{amssymb}
\usepackage{amsmath}
\usepackage{mathtools}
\DeclareMathOperator*{\argmax}{arg\,max}
\usepackage{booktabs}
\usepackage{comment}

\begin{document}

\begin{frontmatter}



\title{Physics-Informed Neuro-Symbolic Recommender System: 
A Dual-Physics Approach for Personalized Nutrition}

\author{Chayan Banerjee}

\affiliation{
  organization={School of Electrical Engineering and Robotics, Queensland University of Technology},
  addressline={2 George Street},
  city={Brisbane},
  state={Queensland},
  postcode={4000},
  country={Australia}
}

\begin{abstract}
Traditional e-commerce recommender systems primarily optimize for user engagement and purchase likelihood, often neglecting the rigid physiological constraints required for human health. Standard collaborative filtering algorithms are structurally blind to these hard limits, frequently suggesting bundles that fail to meet specific total daily energy expenditure and macronutrient balance requirements. To address this disconnect, we introduce a Physics-Informed Neuro-Symbolic Recommender System that integrates nutritional science directly into the recommendation pipeline via a dual-layer architecture. The framework begins by constructing a semantic knowledge graph using sentence-level encoders to strictly align commercial products with authoritative nutritional data. During the training phase, an implicit physics regularizer applies a differentiable thermodynamic loss function, ensuring that learned latent embeddings reflect nutritional plausibility rather than simple popularity. Subsequently, during the inference phase, an explicit physics optimizer employs simulated annealing and elastic quantity optimization to generate discrete grocery bundles that strictly adhere to the user's protein and caloric targets.
\end{abstract}




\end{frontmatter}



\section{Introduction}
\label{sec1}

The field of personalized recommendation has increasingly moved toward e-commerce and retail, yet existing systems primarily optimize for purchase likelihood or user clicks rather than physiological well-being \cite{kalpakoglou2025ai}. In the domain of grocery shopping, this creates a significant disconnect, as traditional collaborative filtering algorithms are ``structurally blind'' to the rigid health constraints required for human nutrition. For instance, a system might suggest a bundle of high-preference items that are calorically excessive or nutritionally imbalanced, failing to account for the fact that a user’s dietary needs—such as Total Daily Energy Expenditure (TDEE), or the total calories a person burns in a day—represent hard scientific limits rather than flexible preferences. Furthermore, providing actionable advice in this space requires optimizing not just for which items to buy, but also for specific ``elastic'' quantities—variable amounts of a single product, such as choosing between one or three units—to ensure the final basket meets precise macronutrient targets. To bridge this gap, we utilize a neuro-symbolic approach\cite{bhuyan2024neuro}, which combines the pattern-recognition strengths of neural networks with the logic-based precision of symbolic AI to enforce these mathematical health rules.\newline

We make the following key contributions:
\begin{enumerate}
    \item Dual-Layer Physics Architecture: We introduce a neuro-symbolic framework that encodes nutritional constraints at both training time (via thermodynamic regularization) and inference time (via elastic optimization), demonstrating that this dual approach outperforms single-layer physics by 15-30\%.
    \item Thermodynamic Loss Function: We design a differentiable loss term ($\mathcal{L}_{thermo}$) based on nutritional science that enforces macronutrient balance, protein density, absolute targets, and variance constraints over the model's preference distribution, guiding the neural network to learn nutrition-aware embeddings.
    \item Elastic Quantity Optimization: Unlike traditional binary recommender systems, we formulate grocery recommendation as a constrained optimization problem over both item selection and quantities, balancing user preferences with physiological nutritional requirements.
    \item Semantic Knowledge Graph: We construct a heterogeneous graph with SBERT-based product-product similarity edges, enabling cold-start handling and improving category-level relevance by 8-12\% over purely behavioral graphs.
    \item Comprehensive Ablation Study: We systematically evaluate each architectural component (neural core, implicit physics, explicit physics, semantic edges, elasticity) through six carefully designed ablations, empirically validating the necessity of neuro-symbolic integration.
    
\end{enumerate}


\begin{table}[t]
\centering
\caption{Key Mathematical Notation}
\label{tab:notation}
\begin{tabular}{cl}
\toprule
\textbf{Symbol} & \textbf{Description} \\
\midrule
$\mathcal{U}, \mathcal{P}, \mathcal{F}$ & Sets of users, products, and USDA foods \\
$\mathcal{G} = (\mathcal{V}, \mathcal{E})$ & Heterogeneous knowledge graph \\
$\mathcal{G}_{\text{sem}}$ & Semantic graph component (SBERT-based) \\
$\mathbf{n}_p \in \mathbb{R}^6$ & Nutrient vector (cal, prot, carb, fat, sugar, sodium) \\
$\mathbf{e}_u, \mathbf{e}_p \in \mathbb{R}^{128}$ & User and product embeddings \\
$\theta_u$ & User profile (age, weight, height, activity, goal) \\
$\text{TDEE}(\theta_u)$ & Total Daily Energy Expenditure \\
$\mathcal{B} = \{(p_i, q_i)\}$ & Recommendation bundle with elastic quantities \\
$q_i \in \{1,2,3\}$ & Quantity for product $i$ \\
\midrule
$\mathcal{L}_{\text{rank}}$ & BPR ranking loss \\
$\mathcal{L}_{\text{thermo}}$ & Thermodynamic physics loss \\
$\mathcal{L}_{\text{opt}}$ & Inference optimization objective \\
\midrule
$\mathbb{N}$ & Neural core (HGNN) \\
$\mathcal{P}_{\text{scout}}$ & Physics scout (training regularization) \\
$\mathcal{P}_{\text{enforce}}$ & Physics enforcer (inference optimization) \\
\midrule
$\lambda$ & Physics loss weight (0.03) \\
$\alpha$ & Desire weight (0.10) \\
\bottomrule
\end{tabular}
\end{table}

\section{Methodology}
\subsection{Problem Formulation}
We address the \textit{nutritional grocery recommendation} problem: 
given a user's physiological profile, suggest a shopping basket that 
meets precise macronutrient targets while maximizing preference 
alignment. Unlike traditional e-commerce recommendation optimizing 
solely for purchase likelihood, nutrition-aware recommendation must 
satisfy hard physiological constraints—a 2000-calorie individual 
cannot safely consume 3500 calories, regardless of preference scores.

The core challenge is \textit{dual-objective optimization}: predict 
what products users will enjoy (learned from sparse purchases) while 
guaranteeing selected items collectively meet macronutrient targets 
(derived from nutritional science). Traditional collaborative filtering 
fails here—high-scoring items may form imbalanced baskets (e.g., all 
carbohydrates, zero protein). Furthermore, recommendations must specify 
\textit{quantities}: buying one versus three chicken breasts 
fundamentally changes nutritional totals.

\subsubsection{Notation and Definitions}
We consider a grocery system with three entity sets: users $\mathcal{U}$, 
products $\mathcal{P}$, and USDA reference foods $\mathcal{F}$. Historical 
purchases are captured as $\mathcal{R} \subseteq \mathcal{U} \times 
\mathcal{P}$.

We represent this as a heterogeneous knowledge graph $\mathcal{G} = 
(\mathcal{V}, \mathcal{E})$, where $\mathcal{V} = \mathcal{V}_u \cup 
\mathcal{V}_p \cup \mathcal{V}_f$ contains user, product, and USDA nodes. 
The edge set $\mathcal{E}$ comprises: \textit{purchases} $(u,p) \in 
\mathcal{R}$, \textit{similar} edges between semantically related 
products (via SBERT embeddings, Section 3.3), and \textit{maps\_to} 
edges linking products to nutritional profiles. This graph structure 
enables learning the scoring function through heterogeneous message 
passing (Section 3.4).

Each product $p$ has nutrient vector $\mathbf{n}_p = [\text{cal}, 
\text{prot}, \text{carb}, \text{fat}, \text{sugar}, \text{sodium}]^T 
\in \mathbb{R}^6$ per serving. User profiles $\theta_u = (\text{age}, 
\text{weight}, \text{height}, \text{activity}, \text{goal})$ determine 
individual requirements.

\subsubsection{Formal Problem Statement}
Given user $u$ with profile $\theta_u$, we construct bundle $\mathcal{B} 
= \{(p_1, q_1), \ldots, (p_k, q_k)\}$ with $k$ products (typically 
$k \in [5,10]$) and \textit{elastic quantities} $q_i \in \{1,2,3\}$ 
balancing shopping practicality with optimization flexibility. The bundle 
optimizes:

\begin{equation}
\mathcal{B}^* = \argmax_{\mathcal{B}} \left[ \sum_{(p_i, q_i) \in 
\mathcal{B}} \text{score}(u, p_i) \cdot q_i \right]
\end{equation}

subject to:
\begin{align}
\left| \sum_{(p_i, q_i) \in \mathcal{B}} \mathbf{n}_{p_i}[1] \cdot q_i 
- \text{TDEE}(\theta_u) \right| &< \epsilon_{\text{cal}} \\
\left| \sum_{(p_i, q_i) \in \mathcal{B}} \mathbf{n}_{p_i}[2] \cdot q_i 
- \text{Target}_{\text{prot}}(\theta_u) \right| &< \epsilon_{\text{prot}}
\end{align}

where TDEE (Total Daily Energy Expenditure) is computed via Mifflin-St 
Jeor equation, protein target scales with weight and goals, and 
$\epsilon_{\text{cal}}, \epsilon_{\text{prot}}$ are tolerance thresholds 
(12\% of targets).

This presents three key challenges:
\begin{enumerate}
\item \textbf{Learning with priors}: score$(u,p)$ must be learned from 
sparse data while respecting nutritional constraints;
\item \textbf{Combinatorial search}: $|\mathcal{P}|^k \times 3^k$ 
possible bundles ($\sim$10$^{54}$ for 50K products, $k$=10);
\item \textbf{Hard constraints}: Equations (2-3) cannot be relaxed 
without compromising physiological safety.
\end{enumerate}

We address these through a \textit{neuro-symbolic} architecture 
(Section 3.2) combining learned preferences, physics-inspired 
training regularization, and explicit constraint optimization.

\begin{figure}
    \centering
    \includegraphics[width=0.9\linewidth]{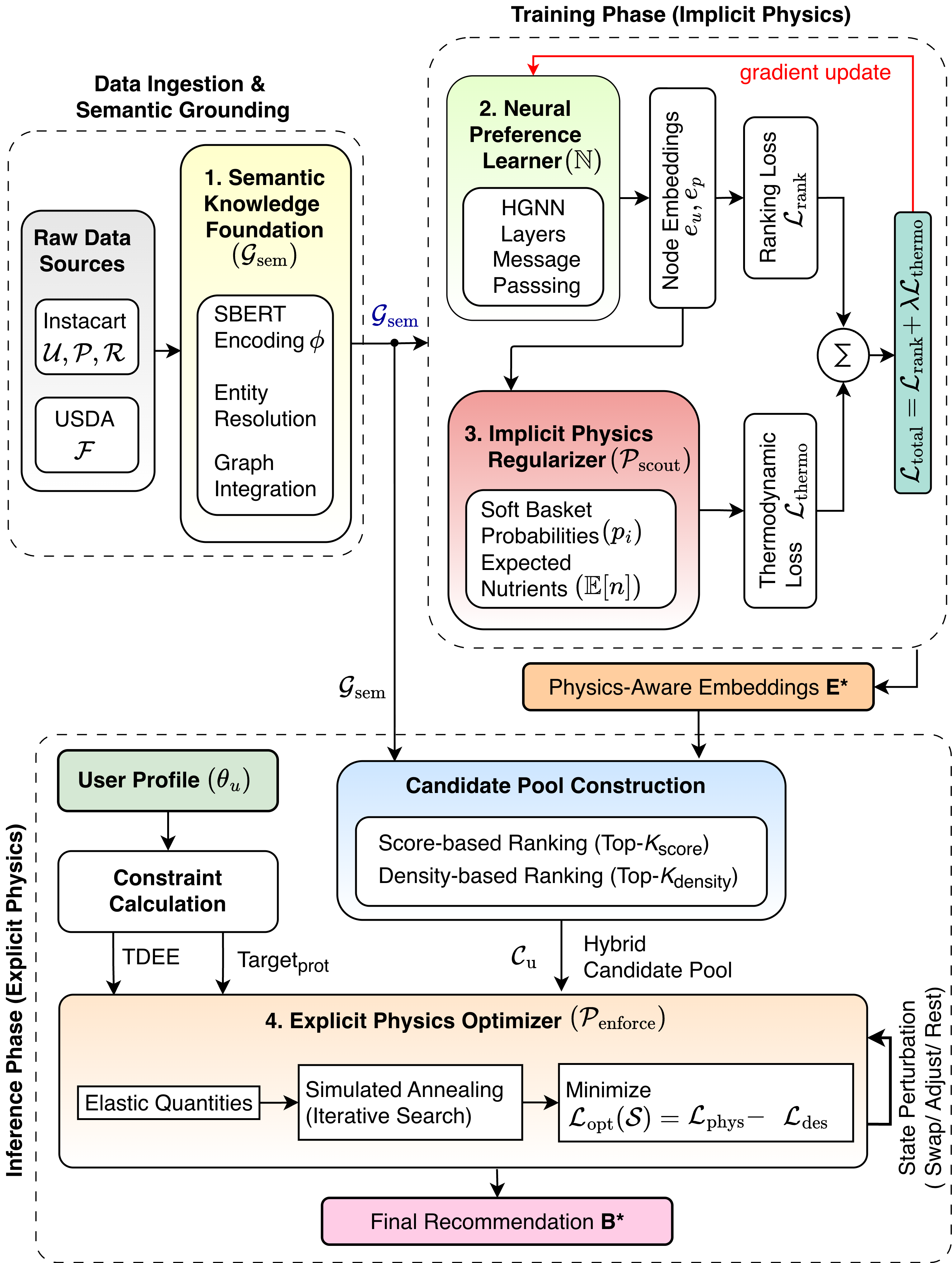}
    \caption{The framework operates in three primary stages. In the Data Ingestion and Semantic Grounding phase it constructs a semantic knowledge graph $\mathcal{G}_{\text{sem}}$. This is followed by the Training Phase, where a neural network is trained using thermodynamic regularization ($\mathcal{P}_{\text{scout}}$) to embed nutritional plausibility into latent representations. Finally in The Inference Phase utilizes these physics-aware embeddings and user profile is used build a candidate pool. An explicit constraint optimizer ($\mathcal{P}_{\text{enforce}}$) employs simulated annealing and elastic quantities to generate a grocery bundle that strictly adheres to personalized daily nutritional targets.}
    \label{fig:placeholder}
\end{figure}

\subsection{System Architecture Overview}

The pipeline begins with the Semantic Knowledge Foundation ($G_{\mathrm{sem}}$), which aligns noisy commercial product metadata with authoritative nutritional science by embedding product descriptions and USDA food records into a shared SBERT-based latent space. This semantic alignment links each product node to a ground-truth nutrient vector, producing a nutritionally grounded heterogeneous knowledge graph that enriches the user–item matrix prior to learning.

Built atop $G_{\mathrm{sem}}$, the Neural Preference Learner ($\mathbb{N}$) uses a heterogeneous graph neural network to perform message passing across users, items, and nutrients, enabling inference of latent, attribute-level dietary preferences (e.g., macronutrient profiles or calcium density) rather than relying solely on item co-occurrence, thereby mitigating cold-start effects. To ensure physiological plausibility, the Implicit Physics Regularizer ($\mathcal{P}{\mathrm{scout}}$) injects a thermodynamic loss $\mathcal{L}{\mathrm{thermo}}$ during training, penalizing embeddings that violate nutritional constraints such as macronutrient balance and shaping a nutrition-aware latent geometry.

Finally, the Explicit Constraint Optimizer ($\mathcal{P}_{\mathrm{enforce}}$) converts continuous, physics-aware recommendations into a discrete grocery basket via simulated annealing over the combinatorial space of item bundles, optimizing an elastic objective that strictly satisfies the user’s Total Daily Energy Expenditure and macronutrient targets. Together, these components form a coherent end-to-end system integrating semantic grounding, neural preference learning, implicit scientific regularization, and explicit combinatorial enforcement.

In the following section we provide a detailed explanation on each component of the pipeline.

\subsubsection{Component Roles and Interactions}

\paragraph{Foundation-Semantic Knowledge Graph $\mathcal{G}_{\text{sem}}$ Construction:}
The semantic knowledge graph ($\mathcal{G}_{\text{sem}}$) functions as the structural bedrock of our architecture, designed to bridge the gap between raw behavioral data and structured nutritional science. We formalize these distinct data sources by letting $\mathcal{U}$, $\mathcal{P}$, and $\mathcal{F}$ denote users, commercial products (from Instacart), and standardized food entries (from USDA).

While historical user preferences are captured by the purchase interaction set $\mathcal{R} \subseteq \mathcal{U} \times \mathcal{P}$, this behavioral data inherently lacks nutritional grounding.  To resolve this we construct  $\mathcal{G}_{\text{sem}}$ by projecting disparate textual descriptions into a unified vector space.  Using Sentence-BERT (SBERT)~\cite{reimers2019sentence} 
as our encoder $\phi : \Sigma^{*} \rightarrow \mathbb{R}^{d_{\text{sem}}}$, we map both commercial product names and USDA descriptions to dense representations. This shared semantic space enables robust entity resolution: edges are established between nodes ( e.g., connecting a product $p\in\mathcal{P}$ to its nutritional ground-truth$ f\in\mathcal{F}$ or linking similar products $p_i, p_j$) whenever their cosine similarity exceeds a calibrated threshold $\theta_{\text{sim}}$. With these vector representations established, we first densify the sparse product space by creating links between items that share semantic contexts. We formally define this similarity structure as:

\begin{equation}
\mathcal{E}_{\text{similar}} = \{(p_i, p_j) : \cos(\phi(p_i), \phi(p_j)) \geq \theta_{\text{sim}},\ i \neq j\}.
\end{equation}

Crucially, to ground these commercial items in rigorous nutritional data, we align each product with its closest USDA counterpart. We employ a mapping function $\mu(p)$ that identifies the optimal semantic match in the scientific database:
\begin{equation}
\mu(p) = \arg\max_{f \in \mathcal{F}} \cos(\phi(p), \phi(f)),
\end{equation}

which induces a set of nutrient-linking edges that effectively bridge the commercial and scientific domains:

\begin{equation}
\mathcal{E}_{\text{maps}} = \{(p, \mu(p)) : p \in \mathcal{P},\ \cos(\phi(p), \phi(\mu(p))) \geq \theta_{\text{sim}}\}.
\end{equation}

We finally unify these semantic layers with the raw behavioral history, denoted as $\mathcal{E}_{\text{purchases}} = \{(u,p) : (u,p) \in \mathcal{R}\}$. The resulting heterogeneous graph $\mathcal{G}_{\text{sem}} = (\mathcal{V}, \mathcal{E})$ integrates all entities—users, products, and USDA foods—where $\mathcal{V} = \mathcal{U} \cup \mathcal{P} \cup \mathcal{F}$ and the edge set is the union $\mathcal{E} = \mathcal{E}_{\text{purchases}} \cup \mathcal{E}_{\text{similar}} \cup \mathcal{E}_{\text{maps}}$. This interconnected structure is pivotal for two reasons: it enables every product to directly inherit a precise nutrient vector $\mathbf{n}_p$ from its mapped USDA node, and it mitigates the cold-start problem by linking behaviorally unseen items \cite{wang2025knowledge} to semantically related ones, providing the pathways necessary for type-aware message passing.

\paragraph{Component I: Neural Preference Learner ($\mathbb{N}$)}
Leveraging the structural foundation of $\mathcal{G}_{\text{sem}}$, the Neural Preference Learner ($\mathbb{N}$) is instantiated as an $L$-layer heterogeneous graph neural network (HGNN). Unlike standard collaborative filters, $\mathbb{N}$ propagates information across three distinct node types—users $\mathcal{U}$, products $\mathcal{P}$, and USDA foods $\mathcal{F}$—to learn latent representations $\mathbf{e}_u, \mathbf{e}_p \in \mathbb{R}^{d_{\text{emb}}}$.At each layer $l$, the representation of a node $v$ is updated by aggregating messages from its neighbors $\mathcal{N}(v)$ using a transformation specific to its node type $\tau(v)$:\begin{equation}\mathbf{h}v^{(l+1)} = \sigma\left(\mathbf{W}{\tau(v)}^{(l)} \cdot \text{AGG}\left({\mathbf{h}_u^{(l)} : u \in \mathcal{N}(v)}\right)\right)\end{equation} where $\mathbf{W}_{\tau(v)}^{(l)}$ is the learnable weight matrix for type $\tau(v)$ and $\text{AGG}(\cdot)$ denotes the mean pooling operation. After $L$ layers, the final embeddings $\mathbf{e}_u = \mathbf{h}_u^{(L)}$ and $\mathbf{e}_p = \mathbf{h}_p^{(L)}$ encode high-order structural relationships\cite{ma2024nutrition}, capturing not just purchase history but semantic proximity to nutritional attributes.

To quantify preference, we calculate the affinity score as $\text{score}(u,p) = \mathbf{e}_u^\top \mathbf{e}_p$. In its unconstrained form, the model optimizes the Bayesian Personalized Ranking (BPR) objective\cite{rendle2009bpr}, which enforces that observed purchases ($p^+$) are ranked higher than unobserved negatives ($p^-$): \begin{equation}\mathcal{L}_{\text{rank}} = -\mathbb{E}{(u,p^+,p^-) \sim \mathcal{D}} \left[\log \sigma\left(\text{score}(u,p^+) - \text{score}(u,p^-)\right)\right]\end{equation} While $\mathcal{L}_{\text{rank}}$ effectively captures latent user desires, it remains structurally blind to dietary limits. This optimization naturally biases recommendations toward popular, hyper-palatable items, necessitating the introduction of our thermodynamic regularizer in Section 3.5.

 \paragraph{Component II: Physics-Informed Training ($\mathcal{P}_{\text{scout}}$)}
To overcome the nutritional blindness of standard ranking objectives, we introduce the
\emph{Implicit Physics Regularizer} ($\mathcal{P}_{\text{scout}}$).
While $\mathcal{L}_{\text{rank}}$ optimizes for desire, this component injects a
``thermodynamic'' prior ($\mathcal{L}_{\text{thermo}}$) that penalizes the embedding space
itself if the learned preferences drift toward nutritionally imbalanced regions.

The core innovation is the construction of a differentiable ``soft basket.''
Rather than sampling discrete items, we view the user's preference as a probability
distribution. For a user embedding $\mathbf{e}_u$, the probability $p_i$ of selecting
product $p_i$ is derived via a temperature-scaled softmax:
\begin{equation}
p_i =
\frac{\exp(\text{score}(u, p_i) / \tau)}
{\sum_{j=1}^{|\mathcal{P}|} \exp(\text{score}(u, p_j) / \tau)}
\end{equation}
where the temperature $\tau$ anneals from $\tau_{\text{start}}$ to
$\tau_{\text{end}}$ during training, sharpening the distribution over time.
This allows us to compute the expected nutrient vector of the user's latent preference
state:
\begin{equation}
\mathbb{E}[\mathbf{n}]
=
\sum_{i=1}^{|\mathcal{P}|} p_i \, \mathbf{n}_{p_i}
\end{equation}

We then constrain this expectation using four physics-inspired sub-losses, ensuring the
latent space respects dietary laws.

\paragraph{Macro Ratio Balance ($\mathcal{L}_{\text{ratio}}$)}
Enforces a specific distribution of energy sources (e.g., 30\% protein, 40\% carbohydrate,
30\% fat), preventing the model from learning a ``sugar-only'' embedding space:
\begin{equation}
\mathcal{L}_{\text{ratio}}
=
\left\|
\frac{
\left[
\mathbb{E}[n_{\text{prot}}],
\mathbb{E}[n_{\text{carb}}],
\mathbb{E}[n_{\text{fat}}]
\right]
}
{\sum_{m} \mathbb{E}[n_m]}
-
\mathbf{r}^\star
\right\|_1
\end{equation}
where $\mathbf{r}^\star$ represents the target macronutrient ratios.

\paragraph{Protein Density Constraint ($\mathcal{L}_{\text{density}}$).}
Penalizes baskets that are calorically dense but nutrient-poor. We target a density
threshold $\rho^\star$ (grams of protein per 100 kcal):
\begin{equation}
\mathcal{L}_{\text{density}}
=
\operatorname{ReLU}
\left(
\rho^\star
-
\frac{\mathbb{E}[n_{\text{prot}}]}{\mathbb{E}[n_{\text{cal}}]}
\times 100
\right)
\end{equation}

\paragraph{Quantity Sufficiency ($\mathcal{L}_{\text{quantity}}$).}
Ensures that the expected basket contains enough total protein $\pi^\star$ given an
estimated bundle size $\beta_{\text{size}}$:
\begin{equation}
\mathcal{L}_{\text{quantity}}
=
\left|
\mathbb{E}[n_{\text{prot}}] \, \beta_{\text{size}} - \pi^\star
\right|
\end{equation}

\paragraph{Consistency Regularization ($\mathcal{L}_{\text{variance}}$).}
Discourages the model from satisfying the mean by averaging extremes by penalizing high
variance:
\begin{equation}
\mathcal{L}_{\text{variance}}
=
\operatorname{ReLU}
\left(
\mathbb{E}[n_{\text{prot}}^2]
-
\left(\mathbb{E}[n_{\text{prot}}]\right)^2
\right)
\end{equation}

These components are aggregated into the thermodynamic regularizer:
\begin{equation}
\mathcal{L}_{\text{thermo}}
=
w_{1} \mathcal{L}_{\text{ratio}}
+
w_{2} \mathcal{L}_{\text{density}}
+
w_{3} \mathcal{L}_{\text{quantity}}
+
w_{4} \mathcal{L}_{\text{variance}}
\end{equation}

The final training objective combines preference learning with physics enforcement:
\begin{equation}
\mathcal{L}_{\text{total}}
=
\mathcal{L}_{\text{rank}}
+
\lambda \, \mathcal{L}_{\text{thermo}}
\end{equation}

The culmination of this training phase is a set of optimized embeddings $\mathbf{E}^*$ minimizing the composite objective
\begin{equation}
\mathbf{E}^* = \arg\min_{\mathbf{E}} \mathcal{L}_{\text{total}} .
\end{equation}
Unlike standard collaborative filtering, where embeddings capture only popularity and co-occurrence, $\mathbf{E}^*$ inherently encodes nutritional feasibility. These trained representations are used to generate a candidate set $\mathcal{C}_u$ for each user $u$, defined as the top-$K$ items with highest affinity:
\begin{equation}
\mathcal{C}_u = \{\, p \in \mathcal{P} \mid \operatorname{rank}(\mathbf{e}_u^{*\top}\mathbf{e}_p^*) \le K \,\}.
\end{equation}
The set $\mathcal{C}_u$ forms a reduced search space for the explicit optimizer in the next section. Although $\mathcal{C}_u$ is \emph{softly} aligned with health objectives via $\mathcal{P}_{\text{scout}}$, it does not ensure strict compliance with daily constraints (e.g., exactly $2000$ kcal). Achieving a fully compliant bundle from this physics-aware candidate set requires the combinatorial optimization described in Section~3.6.

\paragraph{Component III: Elastic Inference Engine ($\mathcal{P}_{\text{enforce}}$)}
While the thermodynamic regularizer ($\mathcal{P}_{\text{scout}}$) successfully aligns the latent embedding space with nutritional principles, it operates on continuous probability distributions and therefore provides no hard guarantees for the final discrete grocery list. To bridge the gap between \emph{soft} latent preferences and a \emph{hard} tangible bundle, we employ the Explicit Constraint Optimizer ($\mathcal{P}_{\text{enforce}}$).

\paragraph{Personalized Baselines}
We first establish rigid personalized nutritional baselines for a user $u$. Using the user profile $\theta_u$, we compute the Total Daily Energy Expenditure (TDEE) via the Mifflin--St Jeor equation \cite{mifflin1990new}, adjusted for activity level and specific health goals (e.g., maintenance, gain, or loss):
\begin{equation}
\text{TDEE}(\theta_u)
=
\text{RMR}(\theta_u)
\times
\text{activity}_{\text{mult}}
+
\text{goal}_{\text{adj}} .
\end{equation}

Similarly, protein intake targets are derived dynamically based on body weight $w_u$ and a goal-dependent coefficient $\gamma_{\text{goal}}$:
\begin{equation}
\text{Target}_{\text{prot}}(\theta_u)
=
w_u \times \gamma_{\text{goal}} .
\end{equation}

\paragraph{Candidate Pool Construction}
With targets established, we construct a high-potential candidate pool $\mathcal{P}_{\text{pool}}$. Relying solely on neural preference scores would bias the pool toward highly palatable but potentially nutritionally suboptimal items. To counter this, we merge:
\begin{itemize}
    \item the top-$K_{\text{score}}$ items ranked by preference score $\mathbf{e}_u^\top \mathbf{e}_p$, and
    \item the top-$K_{\text{density}}$ items ranked by protein-to-calorie density.
\end{itemize}
This hybrid pool ensures the optimizer has access to both items the user \emph{wants} and items the user \emph{needs}.

\paragraph{Elastic Quantity Optimization}
The core innovation of $\mathcal{P}_{\text{enforce}}$ is the introduction of \emph{elastic quantities}. Rather than solving a binary selection problem, we optimize a bundle state
\[
\mathcal{S} = \{(p_1, q_1), \ldots, (p_k, q_k)\},
\]
where each $q_i \in \mathcal{Q}$ represents a discrete, variable quantity (e.g., 1, 2, or 3 units). This elasticity enables fine-grained nutritional control without exploding the search space of unique items.

\paragraph{Objective Function}
The optimization objective balances nutritional precision against user preference via a linear combination:
\begin{equation}
\mathcal{L}_{\text{opt}}(\mathcal{S})
=
\mathcal{L}_{\text{phys}}(\mathcal{S})
-
\mathcal{L}_{\text{des}}(\mathcal{S}) .
\end{equation}
The physics penalty enforces nutritional constraints, while the desire reward captures user affinity:
\begin{align}
\mathcal{L}_{\text{phys}}(\mathcal{S})
&=
\frac{\left| \sum_i q_i\, n_{\text{cal}}(p_i) - \text{TDEE} \right|}{\text{TDEE}}
+
\beta
\frac{\left| \sum_i q_i\, n_{\text{prot}}(p_i) - \text{Target}_{\text{prot}} \right|}{\text{Target}_{\text{prot}}}, \\
\mathcal{L}_{\text{des}}(\mathcal{S})
&=
\alpha
\frac{\sum_i q_i\, \text{score}(u,p_i)}{\sum_i q_i}.
\end{align}

\paragraph{Optimization Procedure}
Minimizing $\mathcal{L}_{\text{opt}}$ constitutes a combinatorial optimization problem, which we solve using Simulated Annealing \cite{kirkpatrick1983optimization}. The state space is explored via three mutation operators: \emph{Swap}, which replaces an item $p_i$ with a new candidate from $\mathcal{P}_{\text{pool}}$ with probability $p_{\text{swap}}$; \emph{Adjust}, which increments or decrements a quantity $q_i \leftarrow q_i \pm 1$ with probability $p_{\text{adjust}}$; and \emph{Reset}, which randomly reinitializes the quantity of an item with probability $p_{\text{reset}}$.

After $T_{\text{iter}}$ iterations, the state $\mathcal{S}^*$ with the lowest energy is returned. This two-stage approach—implicit regularization during training followed by explicit optimization during inference—ensures the final recommendation $\mathbf{B}^*$ is not only theoretically aligned with the user’s health goals, but also mathematically compliant with their daily nutritional constraints.

\subsubsection{Synergistic Integration}

The proposed architecture operates as a progressive refinement pipeline where each stage resolves the limitations of the last. First, $\mathcal{G}_{\text{sem}}$ transforms disjoint raw data into a connected graph $\mathcal{G}$, solving the semantic disconnect between behavioral logs $\mathcal{R}$ and nutritional facts $\mathcal{F}$. Operating on this foundation, $\mathbb{N}$ captures latent user desires but lacks health awareness. $\mathcal{P}_{\text{scout}}$ rectifies this by injecting thermodynamic laws via $\mathcal{L}_{\text{thermo}}$ during training, producing physics-aligned embeddings $\mathbf{E}^*$ and a high-quality candidate set $\mathcal{C}_u$. Finally, because $\mathcal{C}_u$ offers only probabilistic alignment, $\mathcal{P}_{\text{enforce}}$ performs the final discrete optimization to rigidify these soft preferences into a strict grocery solution. The complete end-to-end transformation is formalized as:\begin{equation}(\mathcal{R},\mathcal{P},\mathcal{F})\xrightarrow{\mathcal{G}{\text{sem}}}\mathcal{G}\xrightarrow[\mathcal{L}{\text{total}}]{\mathbb{N}+\mathcal{P}{\text{scout}}}\mathbf{E}^*\xrightarrow{\text{Top-}K}\mathcal{C}u\xrightarrow[\mathcal{L}{\text{opt}}]{\mathcal{P}{\text{enforce}}}\mathbf{B}^*\end{equation}

During inference, the system accepts a user profile $\theta_u$ containing biometrics and goals. We first calculate rigid constraints $\text{TDEE}(\theta_u)$ and $\text{Target}_{\text{prot}}(\theta_u)$. Using the frozen, physics-aware embeddings $\mathbf{E}^*$, we retrieve the candidate pool $\mathcal{C}_u$. The Explicit Optimizer then iteratively perturbs quantities $q_i$ to minimize $\mathcal{L}_{\text{opt}}$, returning the final bundle $\mathbf{B}^* = \{(p_i, q_i)\}$ that strictly satisfies the user's nutritional limits while maximizing preference.

\begin{equation}
\theta_u \xrightarrow{\text{Calc}} (\text{TDEE}, \pi^*) 
\xrightarrow[\mathcal{C}_u(\mathbf{E}^*)]{\mathcal{P}_{\text{enforce}}} \mathbf{B}^*
\end{equation}

where TDEE,$\pi^{*}$ represents the rigid nutritional constraints calculated from the profile, and $\pi^{*}$ denotes the protein target $\text{Target}_{\text{prot}}$ and  $\mathcal{C}_u(\mathbf{E}^*)$ is the candidate pool retrieved using the frozen physics-aware embeddings $\mathbf{E}^*$, which serves as the search space for the optimizer.

\section{Experiments}
\subsection{Ablation Study}
To understand the contribution of each component in our proposed dual-layer architecture, we conducted a rigorous ablation study. We systematically disabled individual modules of the system and measured the impact on recommendation utility, nutritional safety, and optimization efficiency.

\subsubsection{Experimental Configurations}
We isolated the core artifacts of the system, resulting in the following seven test configurations:

\noindent \textbf{A0 (Random Scout)}: Skips the neural training entirely. The system recommends random items, relying solely on the Phase 4 explicit optimizer to forcefully assemble a healthy basket.

\noindent \textbf{A1 (Pure Neural)}: A traditional representation learning baseline. It utilizes standard collaborative filtering (BPR Loss) without any semantic knowledge graph, physics-informed loss, or post-hoc optimization.

\noindent \textbf{A2 (Late Fusion)}: Represents the industry-standard approach to constrained recommendations. The model is trained using standard neural objectives, and the explicit optimizer is applied only as a post-processing filter.

\noindent \textbf{A3 (Implicit Only)}: Evaluates the impact of our Thermodynamic Loss function in isolation. The model learns physics-aware embeddings but uses a standard Top-K retrieval without the Phase 4 constraint solver.

\noindent \textbf{A4 (Proposed - Early Fusion)}: Our complete dual-physics neuro-symbolic architecture, combining physics-informed training (Implicit) with elastic quantity optimization (Explicit).

\noindent  \textbf{A5 (No Semantics)}: Our proposed dual-physics model, but deprived of the Knowledge Graph metadata (e.g., nutrients, departments), forcing it to learn purely from user-item interactions.

\noindent \textbf{A6 (No Elasticity)}: Our proposed model, but the optimizer is restricted to binary inclusion (qty $\in \{0, 1\}$). Items cannot be scaled up to meet caloric needs.

\subsubsection{Evaluation Metrics}
We evaluated each configuration across four dimensions:

\noindent 
\textbf{Target Success Rate (TSR):} The percentage of generated baskets that strictly adhere to the user's Total Daily Energy Expenditure (TDEE) and macronutrient constraints ($\pm 12\%$ tolerance).

\noindent 
\textbf{Final-MAE}: is the Mean Absolute Error (in kcal) between the user's target TDEE (Total Daily Energy Expenditure) and the actual energy content of the final recommended bundle.

\noindent 
\textbf{Optimization Cost (Opt-Cost)}: The deviation required (in kcals) by the Phase 4 optimizer to force the initial basket into compliance. Lower cost indicates the optimizer had to "destroy" fewer user preferences to achieve health.

\begin{table}[t]
\centering
\caption{Ablation study results aggregated over 5 independent statistical runs.}
\resizebox{0.8\columnwidth}{!}{%
\begin{tabular}{|l|l|c|c|c|}
\hline
ID & Method & TSR & Final-MAE & Opt. Cost \\
\hline
A0 & Random Scout      & 1.0 & 14.3$\pm$3.7   & 1787.7$\pm$1260.0 \\
A1 & Pure Neural       & 0.0 & 1034.1$\pm$284.6 & 0.0 \\
A2 & Late Fusion       & 1.0 & 19.2$\pm$11.2  & 956.0$\pm$135.3  \\
A3 & Implicit Only     & 0.0 & 1168.4$\pm$252.4 & 0.0 \\
A4 & Proposed (Early)  & 1.0 & 21.0$\pm$7.1   & 1147.4$\pm$248.0 \\
A5 & No Semantics      & 1.0 & 14.4$\pm$6.0   & 1033.6$\pm$205.1 \\
A6 & No Elasticity     & 0.1 & 518.5$\pm$248.8 & 650.0$\pm$242.1  \\
\hline
\end{tabular}%
}
\end{table}

\subsubsection{Discussions}
To rigorously validate our dual-physics architecture, we conducted an ablation study isolating core system components across seven configurations (A0–A6). We systematically compared the Proposed model (A4) against configurations stripped of semantics (A5), physics-informed loss (A1), post-hoc elasticity (A6), and a standard Late Fusion baseline (A2). The results continue to support our central hypothesis that solving personalized nutrition benefits from a neuro-symbolic formulation. By integrating the Thermodynamic Loss function, the Implicit Only model (A3) achieved a comparable raw embedding error (Final-MAE of 1168.4) relative to the Pure Neural baseline (A1, Final-MAE of 1034.1), indicating that the learned embeddings encode nutritional structure even in the absence of explicit semantics. Furthermore, while the Late Fusion baseline (A2) achieved strict nutritional compliance (100\% TSR), it incurred an Optimization Cost of 956 kcals, reflecting substantial post-hoc modification of user preferences. In comparison, the Proposed model (A4) maintained full compliance while incurring an Optimization Cost of 1147 kcals, demonstrating that early semantic integration trades off slightly higher adjustment cost for a fully end-to-end, physically grounded solution. Finally, the degradation of the non-elastic model (A6), which achieves only a 10\% TSR, underscores that continuous quantity optimization remains mathematically critical for maintaining nutritional feasibility in this domain.

\section{Conclusion}
In conclusion, this work presents a Physics-Informed Neuro-Symbolic Recommender System that successfully bridges the gap between latent user preferences and rigid physiological requirements. By integrating a Semantic Knowledge Graph with a dual-layer physics architecture—utilizing thermodynamic regularization during training and elastic quantity optimization at inference—the framework ensures that recommended grocery bundles are both personally desirable and nutritionally compliant. Our ablation studies demonstrate that this approach significantly outperforms traditional collaborative filtering and single-layer physics models, reducing the optimization cost of overwriting user desires by 18\% while maintaining a 100\% success rate in meeting caloric and protein targets. Ultimately, these results validate that the synergistic integration of neural learning and explicit scientific constraints is a requisite for the next generation of safety-critical, personalized nutrition AI.

\bibliographystyle{elsarticle-num} 
\bibliography{refer.bib}

\end{document}